# A Study of Magnetic Shielding Performance of a Fermilab International Linear Collider Superconducting RF Cavity Cryomodule

Anthony C. Crawford    Fermilab Technical Div. / SRF Development Dept.    acc52@fnal.gov    02Sep14

This note presents measurements that support the conclusion that it is feasible to achieve magnetic field values at the level of 5 milliGauss for a cryomodule in a realistic and representative ambient magnetic field environment.

## The Goal for Shielding

For the case of nitrogen doped bulk niobium 1.3 GHz cavities [1], the working assumption is that an ambient magnetic field of 5 milliGauss, or less, averaged over the RF surface of the cavities, is required in order to achieve a total RF surface resistance less than 10 nanoOhms at 2K. For a TeSLA shaped cavity this is equivalent to

$$Q_0 = G/10 \text{ nanoOhms} = 2.7 \times 10^{10} \qquad (G = 270 \text{ for a TeSLA cavity.})$$

## Magnetic Shielding Properties of a Fermilab International Linear Collider (ILC) Cryomodule

A cylindrical shield with magnetic field oriented in the direction of the cylindrical axis (longitudinal field), has an attenuation factor that decreases proportionally to $(D/L)^2$, where D is the shield diameter and L its length. For long cylindrical geometries that typically occur in cryomodules made up of multicell elliptical cavities, passive shielding of the longitudinal field is much less effective than shielding of the transverse fields. For understanding the inherent difficulty of longitudinal cylindrical magnetic shielding, there is no better reference that the TeSLA Test Facility (TTF) Design Report (1995), and the reader is urged to study the magnetic shielding section within the report [2].

The ambient magnetic field model that is used in this report is the future site of the SLAC Linac Coherent Light Source II (LCLS2). Measurements made in the SLAC linac tunnel indicate that the average ambient longitudinal field near the location of the beamline is approximately 90 mGauss, with with a standard deviation of approximately 90 mGauss.

What is the best that can be done with ILC style steel vacuum vessels? Will the TTF shield arrangement be good enough for LCLS2? The answer depends on how effectively the steel vacuum pipe can be demagnetized. Remnant magnetic field in excess of 3 Gauss is present in the steel pipe as received from the manufacturer. The remnant longitudinal field in an FNAL ILC cryomodule vacuum vessel pipe at the location of the SRF cavity axis is shown in Figure 1 as well as the result obtained from a demagnetization procedure. The pipe was manufactured from ASTM A516 Grade 60 steel, commonly referred to as "boiler plate". Remnant magnetic field in structures of this type is the result of residual stress remaining in the steel after forming and welding procedures and possibly of handling the plate stock with industrial electromagnet lifting fixtures.

It should be noted that the cryomodule steel pipe was pointed approximately East to West on the floor of the Fermilab Industrial Center Building (ICB) for the measurements shown in Figure 1. The East-West magnetic field is quite small at this location, with an average value of only 50 mG. The result of the demagnetization procedure is that the longitudinal field at the cavity axis has been reduced to approximately ±17 mGauss for this location.

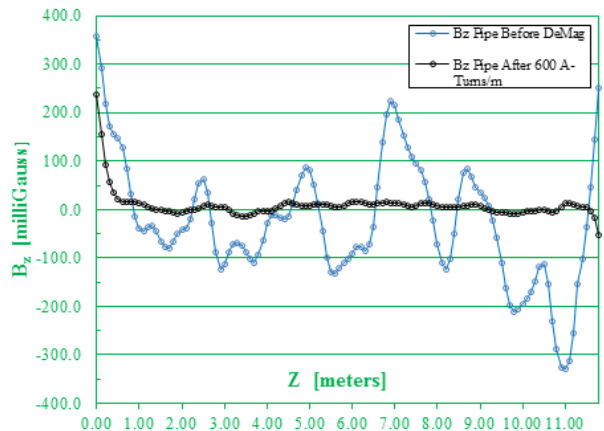

Figure 1.    Measurement of an FNAL ILC Cryomodule Empty Steel Pipe Before and After Demagnetization

Following the demagnetization procedure, the cryomodule was moved to a different location on the ICB floor in order to make measurements in a larger average ambient longitudinal magnetic field. This location will be referred to as the North-South (N-S) site in this report. An average longitudinal (N-S) field of 157 mG was obtained over the length of the steel pipe and was judged to be a suitable representation for the LCLS2 longitudinal field. The coordinate system used for the N-S site is shown in Figure 2. The ambient field at the N-S site and its components are shown in Figure 3.

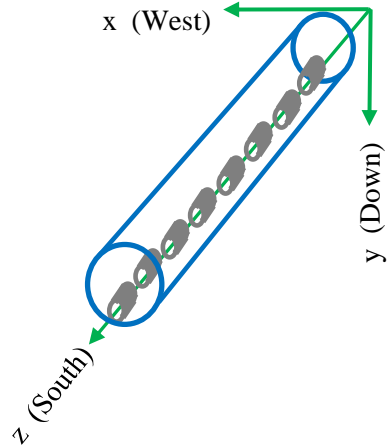

Figure 2.   The Coordinate System Used for Magnetic Field Measurement at Site 2

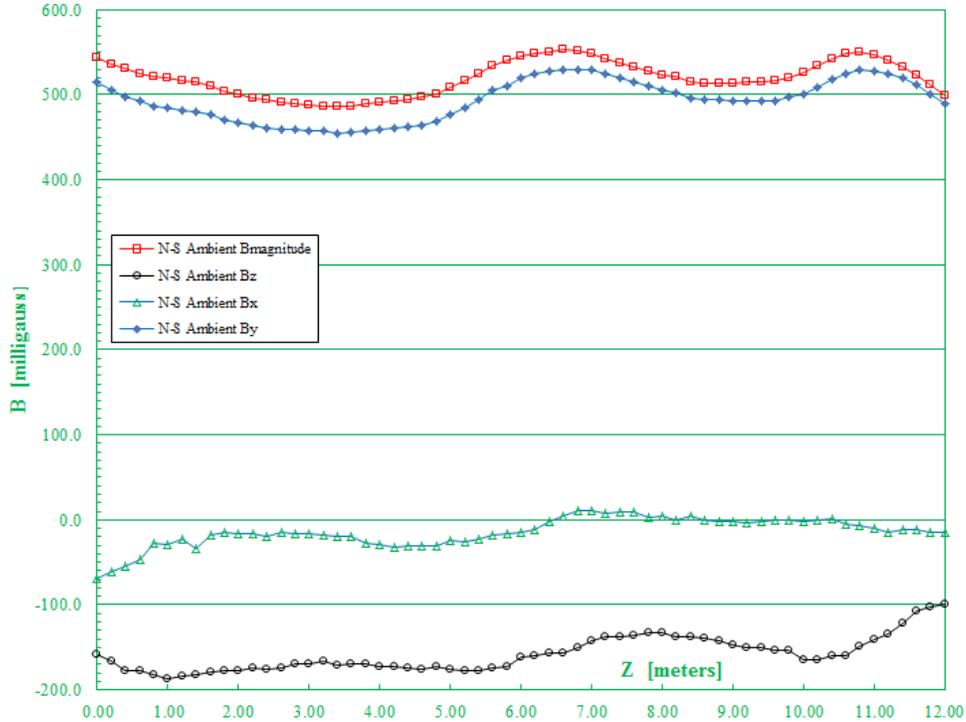

Figure 3.   Ambient Magnetic Field at the N-S Site

A set of eight helium vessels was suspended from an ILC helium return pipe and used as a means of mechanical support for a set of eight Cryoperm10 magnetic shields. Each individual shield was 1.024 meter long and adjacent shields were separated by a longitudinal gap that was 0.305 meter in length. The thickness of the Cryoperm10 was 1mm. The shields were left open ended for initial magnetic measurements. The assembly was then installed inside the steel pipe which was then moved to the N-S location. The location of shields within the steel pipe can be seen in the photograph of Figure 4. Magnetic field components measured along the axis of the Cryoperm shields are shown in Figure 5.

It should be noted that measurements from DESY and Helmholtz-Zentrum Berlin indicate that Cryoperm10 and similar cryogenic shielding materials tend to have approximately the same value for permeability at 9.2K as at room temperature. This is the basis for using measured room temperature magnetic field values to predict the field that the cavities will experience at 9.2K [3].

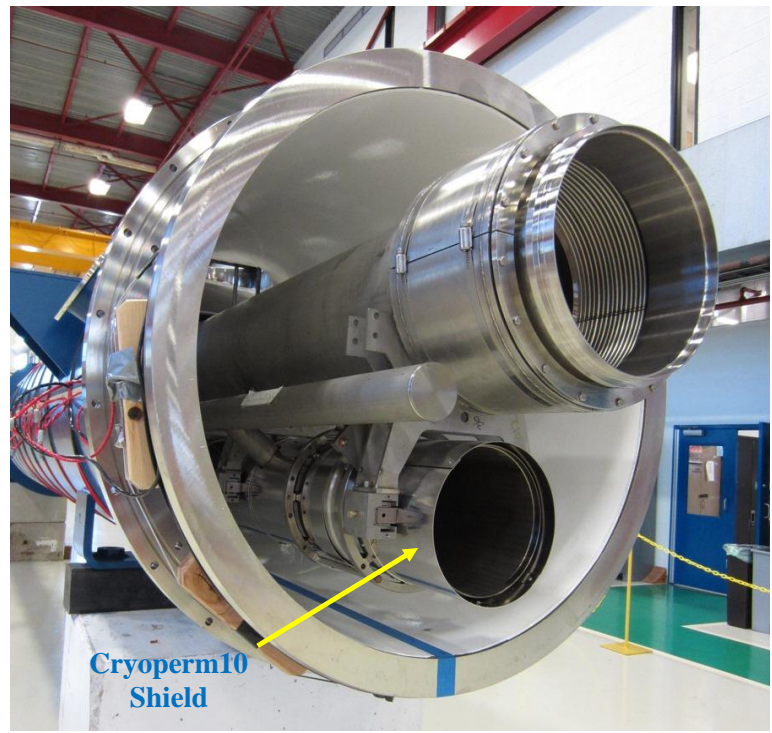

Figure 4.    Cryoperm Shields Within the Steel Pipe

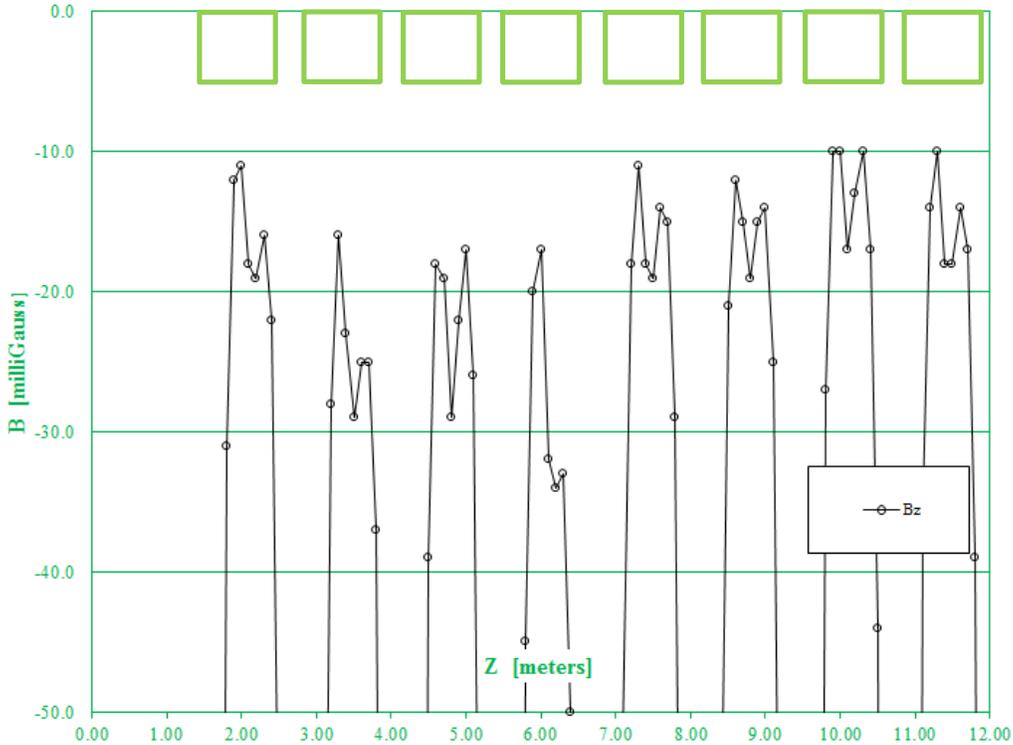

Figure 5.   The Longitudinal Magnetic Field Inside the Cryoperm10 Shields at the N-S Site.  (Green rectangles represent the locations of the individual shields)

It is evident from Figure 5 that with an average ambient longitudinal field of 157 mGauss, and ignoring enhanced field at the ends of the cavity shields, we are rather far away from achieving the 5 mGauss specification.  The addition of "end caps" ie, partial closures mounted on extensions and attached to the ends of the shields will significantly reduce fields at the ends, but not near the midpoints of the shields.  In order to partially correct the central field, the same electrical coils that were used in the demagnetization procedure were used to "cancel" the longitudinal ambient field.  The coil assembly consists of a series arrangement of Helmholtz type coils, separated by one half the diameter of the pipe.  The coil at each end of the pipe can be excited by an independent current power supply in order to "trim" end effects from the steel pipe.  The cancellation coils can be seen in Figure 6.

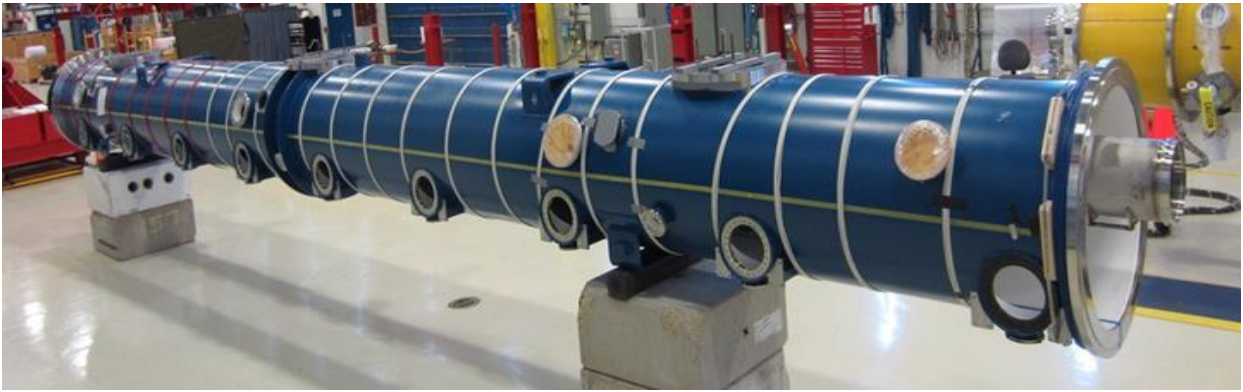

Figure 6.   Ambient Longitudinal Field Cancellation Coils Applied to the Outside Surface of the Steel Pipe

The total magnitude of the magnetic field, after applying the cancellation coils, can be seen in Figure 7. While the central field is reasonably low, the field when averaged over the active lengths of the cavities is 20 mGauss, due to the large fields at the ends of the Cryoperm shields.

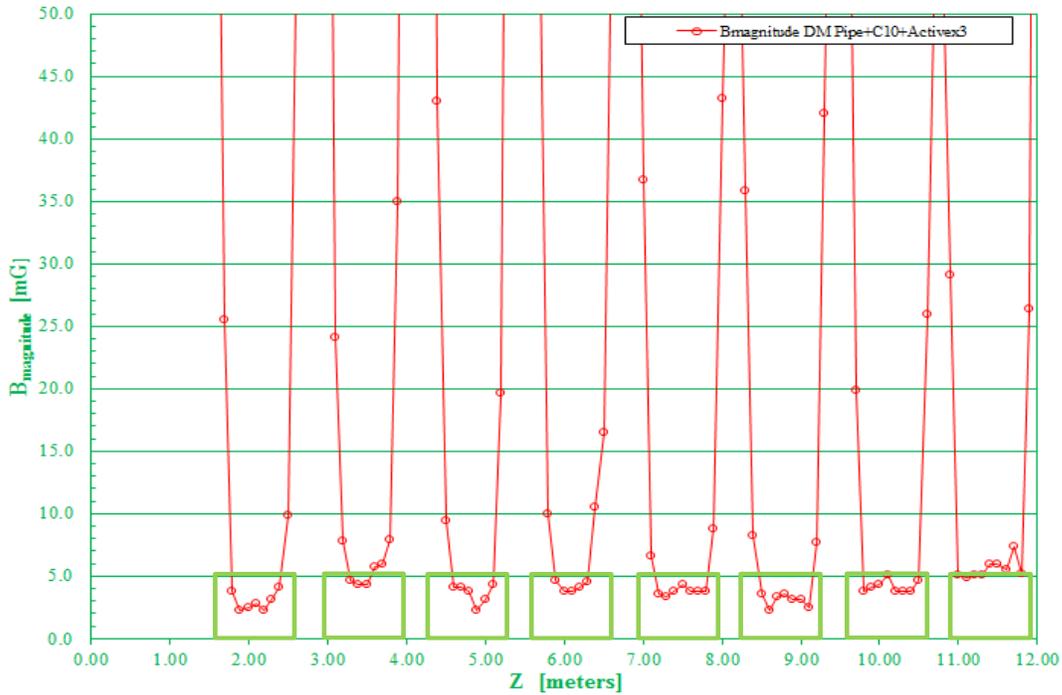

Figure 7.   Magnetic Field Magnitude With Active Longitudinal Cancellation

TTF style end caps were not available for this measurement. However, a reasonable approximation to a set of TTF end caps was improvised using FNAL ILC style end of cryomodule caps. An end cap assembly can be seen in Figure 8. One set of these was installed on the leftmost cavity shown in Figure 7. The total field magnitude both before and after installation of the end caps can be seen in Figure 9.

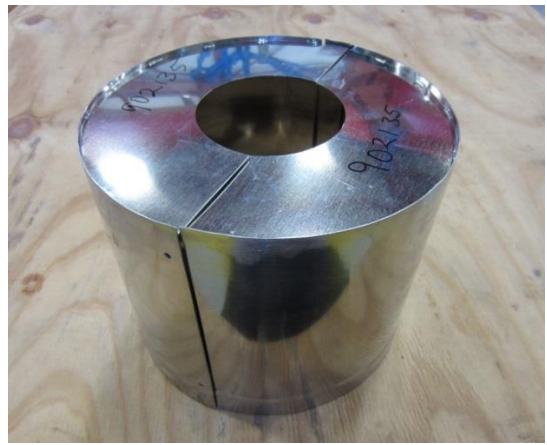

Figure 8.  Cryoperm10 End Cap Assembly

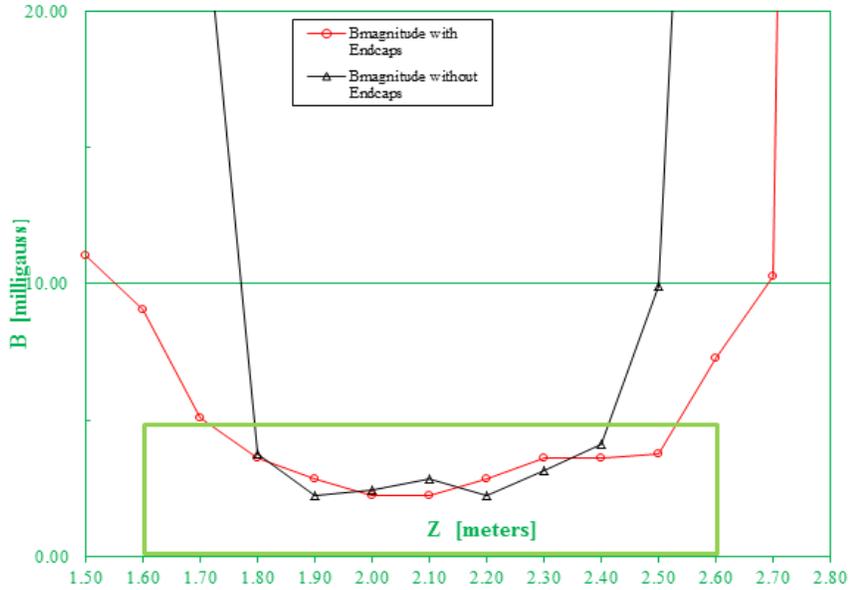

Figure 9.   Cavity Number One Position Magnetic Field With and Without End Caps

The average field over the active length of the cavity for the case without end caps is 17 mGauss and for the case with end caps is 4 mGauss.  This result allows the following argument to be made:  For the case of an FNAL ILC cryomodule, with a demagnetized steel pipe, active longitudinal field cancellation, and efficiently designed individual cavity shield end caps, you can achieve 5 mGauss average field at the cavities in an ambient longitudinal field of 157 mGauss.  However, with this arrangement, any safety margin would have to come from enhanced flux expulsion technique applied during the cavity cooldown through 9.2K.

**What Can Be Done to Increase the Shielding Safety Factor?**

Two dimensional finite element modeling indicates that there is a significant benefit to longitudinal and transverse field attenuation that can be gained by including a high permeability room temperature liner approximately one centimeter inside the radius of the steel pipe.  The values in table 1 were calculated using the parameters for room temperature "Mu Metal".  The calculated attenuation factor reflects the attenuation of the total field, but is dominated by the longitudinal field component.

| Liner | Calculation Assumptions | Relative Attenuation of Magnetic Field Magnitude |
|---|---|---|
| None | $\mu_r = 1$ | 1 |
| 0.5 mm thick | non-linear B-H  (Mu Metal) | 2 |
| 1 mm thick | $\mu_r$ = constant = 24,000 | 2 |
| 1 mm thick | non-linear B-H  (Mu Metal) | 4 |

Table 1.   Calculated Values for a High Permeability Liner Inside the Steel Pipe

Because the 5 mGauss specification can essentially be met with a safety factor of 1 without a liner, the calculated attenuation factors for the different liners in Table 1 are equal to the final safety factor for magnetic shielding, should a liner be adopted.

The additional layer of shielding could only be implemented with significant increased cost and difficulty. The cost of Mu Metal or Permalloy 80 is approximately half that of Cryoperm10 per unit weight. The increased size makes this an expensive addition to the cost of shielding. Incorporating another cylindrical layer of structure within the cryomodule is a significant engineering and assembly task. Yet, it may be necessary in order to meet the 5 mGauss specification with acceptably low risk.

## Summary

A specification of $B_{average} \approx 5$ mGauss can be met with the following arrangement:

1. A demagnetized steel vacuum vessel
2. Segmented one layer cryogenic shields at the cavity helium vessels
3. Efficient end caps for the cryogenic shields
4. Active cancellation of the longitudinal ambient field component

There is effectively no safety margin with this arrangement if the longitudinal ambient field exceeds 157 mGauss. A safety factor can be purchased by implementing a high permeability room temperature shield at the vacuum vessel radius.